\documentclass[fleqn,10pt]{article}
\usepackage{epsfig}
\newcommand{\be}{\begin{equation}}
\newcommand{\ee}{\end{equation}}

\newcommand{\eq}[1]{Eq.\,(\ref{#1})}

\begin{document}
\setcounter{secnumdepth}{4}
\renewcommand\thepage{\ }
%
%
\begin{titlepage} 
%
\newcommand\reportnumber{989} 
\newcommand\mydate{May 18, 2004} 
\newlength{\nulogo} 
\settowidth{\nulogo}{\small\sf{UW Report  MADPH-04-XXXX}}
\title{
\vspace{-.8in} 
\hfill\fbox{{\parbox{\nulogo}{\small\sf{
NUHEP Report  \reportnumber\\
UW Report MADPH-04-1382\\
          \mydate}}}}
\vspace{0.5in} \\
{
Evidence for the saturation of the Froissart bound 
}}

\author{
M.~M.~Block\\
{\small\em Department of Physics and Astronomy,} \vspace{-5pt} \\ 
{\small\em Northwestern University, Evanston, IL 60208}\\
\vspace{-5pt}
\  \\
F.~Halzen
\vspace{-5pt} \\ 
{\small\em Department of Physics,} 
\vspace{-5pt} \\ 
{\small\em University of
Wisconsin, Madison, WI 53706} \\
\vspace{-5pt}\\
%
\vspace{-5pt}\\
%
}    
\vspace{.5in}
\vfill
\date {}
\maketitle
\begin{abstract}
It is well known that fits to high energy data cannot discriminate between asymptotic $\ln s$ and $\ln^2s$ behavior of total cross section. We show that this is no longer the case when we impose the condition that the amplitudes also describe, on average, low energy data dominated by resonances. We demonstrate this by fitting real analytic amplitudes to high energy measurements of the $\gamma p$ total cross section, for $\sqrt s\ge 4$~GeV. We subsequently require that the asymptotic fit smoothly  join the $\sqrt s=$2.01~GeV  cross section described by Dameshek and Gilman as a sum of Breit-Wigner resonances. The results strongly favor the high energy $\ln^2s$ fit of the form $\sigma_{\gamma p}=c_0 +c_1{\ln }(\nu/m)+c_2{\ln }^2(\nu/m)+\beta_{\cal P'}/\sqrt{\nu/m}$, basically excluding a $\ln s$ fit of the form $\sigma_{\gamma p}=c_0 +c_1{\ln }(\nu/m)+\beta_{\cal P'}/\sqrt{\nu/m}$, where $\nu$ is the laboratory photon energy. This evidence for saturation of the Froissart bound for $\gamma p$ interactions is confirmed by applying the same analysis to  $\pi p$ data using vector meson dominance.
\end{abstract}
\end{titlepage} 
%
\pagenumbering{arabic}
\renewcommand{\thepage}{-- \arabic{page}\ --}  

The Froissart bound\cite{froissart} states that the high energy cross section for the scattering of hadrons is bounded by $\sigma =2\pi(1/\mu)^2\ln^2(s/s_0)$, where $s_0$ is a scale factor and $\mu$ the pion mass. This fundamental result is derived from unitarity and analyticity. Clearly, the coefficient $2\pi(1/\mu)^2$ of $\approx 16,000$~mb, for $\sqrt s=2000$~GeV, assuming an $s_0$ of 50~GeV$^2$, is much larger than  the measured $\bar p p$ cross section of order $\approx$ 80~mb. In this context, saturating the Froissart bound refers to an energy dependence of the total  cross section rising no more rapidly than  $\ln^2s$.

The question as to whether any of the present day  high energy data for  $\bar pp$, $pp$, $\pi^+ p$, $\pi^-p$, $\gamma p$ and $\gamma \gamma$ cross sections saturate the Froissart bound has not been settled; one can not discriminate between asymptotic fits of $\ln s$ and $\ln^2 s$  using high energy data only\cite{bkw,cudell}.  We here point out that this ambiguity is resolved by requiring that the fits to the high energy data smoothly join the cross section and energy dependence obtained by averaging the resonances at low energy. Imposing this duality\cite{igi} condition, we show that only fits to the high energy data behaving as $\ln^2 s$  smoothly join to the low energy data  on $\gamma p$ total cross sections in the resonance region. 

\begin{figure}[h,t,b] 
\begin{center}
\mbox{\epsfig{file=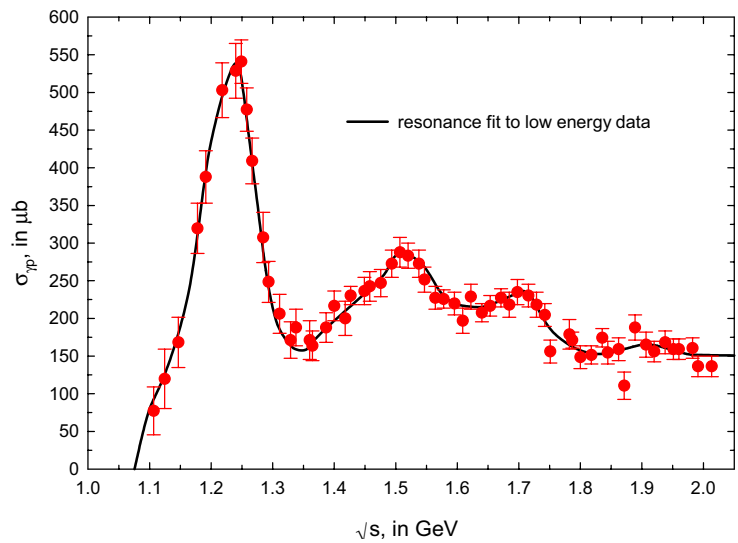
,width=4.4in%
            ,
		bbllx=60pt,bblly=250pt,bburx=530pt,bbury=585pt,clip=%
}}
\end{center}
\caption[]{ \footnotesize
Shown is a fit by Damshek and Gilman\cite{gilman} of the low energy $\sigma_{\gamma p}$ data to a sum of five Breit-Wigner resonances plus a sixth-order polynomial background. The fitted value of $\sigma_{\gamma p}$ at $\sqrt s=2.01$ GeV is 151 $\mu$b.
}
\label{fig:siggpresonances}
\end{figure}

For describing the low energy cross sections we use a convenient parametrization by Damashek and Gilman\cite{gilman} of the forward Compton scattering amplitudes, which yields a very accurate description of the low energy data. It provides us with a best fit in the energy region $2m\nu_0+m^2\le \sqrt s\le 2.01$~GeV using five Breit-Wigner resonances and a 6th order polynomial in $(\sqrt s -\sqrt s_{\rm threshold})$.  Here $\nu_0=m_\pi + m_\pi^2/m$ is the threshold and $m$ the proton mass. Their result is shown in Fig.\,\ref{fig:siggpresonances}.

Following Block and Cahn\cite{bc}, we write the crossing-even real analytic amplitude for high energy $\gamma p$ scattering as\cite{compton}
\begin{equation}
f_+=i\frac{\nu}{4\pi}\left\{A+\beta[\ln (s/s_0) -i\pi/2]^2+cs^{\mu-1}e^{i\pi(1-\mu)/2}-i\frac{4\pi}{\nu}f_+(0)\right\},\label{evenamplitude_gp}
\end{equation}
where $A$, $\beta$, $c$, $s_0$ and $\mu$ are real constants. The variable $s$ is the square of the c.m. system energy and $\nu$ the laboratory momentum. The additional real constant $f_+(0)$ is the subtraction constant at $\nu=0$ introduced in the singly-subtracted dispersion relation\cite{gilman} for the reaction $\gamma +p\rightarrow\gamma + p$. It is fixed in the Thompson scattering limit $f_+(0)=-\alpha/m=-3.03\  \mu {\rm b\  GeV}$. Using the optical theorem, we obtain the total cross section
\be
\sigma_{\gamma p}= A+\beta\left[\ln^2 s/s_0-\frac{\pi^2}{4}\right]+c\,\sin(\pi\mu/2)s^{\mu-1}  \label{sigmatot}
\ee
and  $\rho$, the ratio of the real to the imaginary part of the forward scattering amplitude,  given by
\be
\rho={1\over\sigma_{\rm tot}}\left\{\beta\,\pi\ln s/s_0-c\,\cos(\pi\mu/2)s^{\mu-1}+\frac{4\pi}{\nu} f_+(0)\right\}.\label{rhogeneral}
\ee 
Introducing the definitions $A = c_0 + (\pi^2/4)c_2 - (c_1^2 /4c_2)$, $ s_0 = 2m ^ 2 e^{-c_1 / (2c_2)}$, $\beta=c_2$ and\break $c = (2m^2)^{1 - \mu} / \sin(\pi\mu/ 2) \beta_{\cal P'}$, 
\eq{sigmatot} and \eq{rhogeneral} can, in the high energy limit $s\rightarrow2m\nu$, be written as
\begin{eqnarray}
\sigma_{\gamma p}&=&c_0+c_1\ln\left(\frac{\nu}{m}\right)+c_2\ln^2\left(\frac{\nu}{m}\right)+\beta_{\cal P'}\left(\frac{\nu}{m}\right)^{\mu -1},\label{sigma}\\
\rho_{\gamma p}&=& {1\over\sigma} \left\{{\pi}{2}c_1+\pi c_2\ln\left(\frac{\nu}{m}\right)-\cot(\pi\mu/2)\beta_{\cal P'})\left(\frac{\nu}{m}\right)^{\mu -1}+\frac{4\pi}{\nu}f_+(0) \right\}\label{rho}.
\end{eqnarray}
This transformation linearizes  \eq{sigma} in the real  coefficients $c_0,c_1,c_2$ and $\beta_{\cal P'}$, convenient for a straightforward $\chi^2$ fit to the experimental  $\gamma p$ total cross sections.  Throughout we will use units of $\nu$ in GeV and cross section in $\mu$b. 

Our strategy is to constrain the high energy fit with the precise low energy fit at $\sqrt s\le 2.01$ GeV, which is the energy where Damashek and Gilman\cite{gilman}  join the energy region dominated by resonances to a Regge fit, $a +b/\sqrt{\nu/m}$. They find that the cross section at $\sqrt s=2.01$ is 151 $\mu$b and the slope $d\sigma_{\gamma p}/d(\nu/m)$ is $-b/(\nu/m)^{1.5}$, or $-15.66$ in $\mu$b units. Using the asymptotic asymptotic expression of \eq{sigma}, we thus obtain two constraints
\begin{eqnarray}
\beta_{\cal P'}&=&73.0+2.68c_1 +3.14c_2\label{deriv},\\
 c_0&=&151-0.586c_1-0.343c_2-0.746\beta_{\cal P'},\label{intercept}
\end{eqnarray}
by matching the values of the slope and the cross section, respectively. Unless stated otherwise, both constraints are used in our $\chi^2$ fitting procedure.  

\begin{figure}[h,t,b] 
\begin{center}
\mbox{\epsfig{file=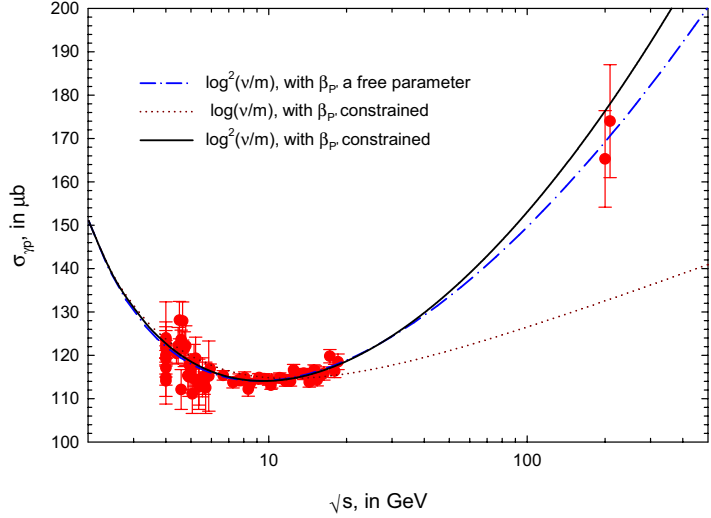,width=4.4in%
            ,
		bbllx=60pt,bblly=170pt,bburx=550pt,bbury=585pt,clip=%
}}
\end{center}
\caption[]{ \footnotesize
 The fitted $\sigma_{\gamma p}$, in $\mu$b, {\em vs.} $\sqrt s$, in GeV. 
The solid curve (Fit~1)  is a $\chi^2$ fit to  the high energy data  of the form $\sigma_{\gamma p}=c_0 +c_1{\ln }(\nu/m)+c_2{\ln }^2(\nu/m)+\beta_{\cal P'}/\sqrt{\nu/m}$, with both $c_0$ and $\beta_{\cal P'}$ constrained by \eq{deriv} and \eq{intercept}. The constaints were derived from a low energy fit to the resonance region and  a Regge $\cal P'$ trajectory\cite{gilman}.  The dot-dashed line is a ${\rm log}^2(\nu/m)$ fit (Fit~2) that constrains $c_0$ only, allowing $\beta_{\cal P'}$ to be a free parameter.  
The dotted line (Fit~3) uses $\sigma_{\gamma p}=c_0 +c_1{\ln }(\nu/m)+\beta_{\cal P'}/\sqrt{\nu/m}$, with both $c_0$ and $\beta_{\cal P'}$ constrained by \eq{deriv} and \eq{intercept}.  The laboratory energy of the photon is $\nu$ and $m$ is the proton mass.  The data used in all fits are the cross sections with $\sqrt s \ge 4$ GeV. All fits pass through the low energy anchor point at $\sqrt s=2.01$~GeV, where $\sigma_{\gamma p}=151\,u$b. Fits 1 and 3 are further constrained to have the same slope as the low energy fit, at $\sqrt s=2.01$~GeV.  Details of the 3 fits are given in Table \ref{table:fits}.
}
\label{fig:siggplogsq+log}
\end{figure}

We next fit  the asymptotic form of \eq{sigma} to the high energy data in the energy range $4\le\sqrt s\le 210$~GeV.  The lower energy data are from the Particle Data Group\cite{pdg}; the high energy points at $\sqrt s=200$ and $\sqrt s=209$~GeV are from the H1 collaboration\cite{h1} and Zeus\cite{zeus} collaborations, respectively. The results are summarized in Table~\ref{table:fits}. For Fit~1,  the data are fitted with a $\ln^2(\nu/m)$ energy dependence imposing constraints \eq{deriv} and \eq{intercept}. We thus obtain fitted values for $c_1$ and $c_2$, which then determine $c_0$ and $\beta_{\cal P'}$. The fit is excellent, yielding a total $\chi^2$ of 50.34 for 61 degrees of freedom, with a fit probability of 0.83.  The fit is shown as the solid line in Fig.\,\ref{fig:siggplogsq+log}. In order to verify that the data discriminates between a $\ln^2(\nu/m)$ fit and a $\ln(\nu/m)$ fit, we made Fit~3 assuming a $\ln(\nu/m)$ energy dependence, {\em i.e.,}\ $c_2=0$. After fitting $c_1$, we determine $c_0$ and $\beta_{\cal P'}$ from the constraint equations. The fit is poor with a total $\chi^2$ of 102.8 for 62 degrees of freedom.  This corresponds to a chance probability of $8.76\times10^{-4}$. It is plotted as the dotted line in Fig.\,\ref{fig:siggplogsq+log} and clearly underestimates the high energy cross measurements. Finally, to test the stability of the $\ln^2(\nu/m)$ fit, we relax the condition that the slopes of the low energy fit and the asymptotic fit are the same at $\sqrt s=2.01$ GeV and only impose the cross section constraint of \eq{intercept}.  Thus, in Fit~2, we fit $c_1,\ c_2$, and $\beta_{\cal P'}$, which then determines $c_0$. This also yields a good fit, with a total $\chi^2$ of 47.48, for 60 degrees of freedom, corresponding to a chance probability of 0.88. Fit~2 is shown as the dashed-dotted line in Fig.\,\ref{fig:siggplogsq+log}. It fits the data well, indicating stability of the procedure. Clearly, the constraints imposed by the low energy data strongly restricts the asymptotic behavior\cite{sensitivity}. 

\begin{table}[h,t]                   
%
\def\arraystretch{1.5}            
\begin{center}
\begin{tabular}[b]{|l||c|c||c||}
     \cline{2-4}
      \multicolumn{1}{c|}{}
      &\multicolumn{2}{c||}{$\sigma \sim \log^2(\nu/m)$}
      &\multicolumn{1}{c|}{$\sigma \sim \log(\nu/m)$}\\
      \hline
      Parameters&Fit 1 &Fit 2&Fit 3 \\ 
	&$c_0$ and $\beta_{\cal P'}$ constrained&$c_0$ constrained&$c_0$ and 		$\beta_{\cal P'}$ constrained\\
	\hline
     $c_0$$\ $ $(\mu$b)&105.64&92.5&84.22 \\
      $c_1$$\ $  $(\mu$b)&$-4.74\pm1.17$&$-0.46\pm 2.88$&$4.76\pm0.11$  \\ 
	$c_2$$\ $  $(\mu$b)&$1.17\pm0.16$&$0.803\pm 0.273$&-----   \\
      $\beta_{\cal P'}$ $(\mu$b)&$64.0$&$78.4\pm9.1$&$85.8$   \\ 
	$\mu$ &0.5&0.5&0.5\\\hline
     	$\chi^2$&50.34&47.48&102.8\\
	d.f.&61&60&62\\
	$\chi^2/$d.f.&0.825&0.791&1.657\\
Probability&0.83&0.88&$8.76\times10^{-4}$\\
     \hline
\multicolumn{4}{c}{$m$ is the proton mass and $\nu$ is the laboratory photon energy}
\end{tabular}
     \caption{\protect\small The fitted results.\label{table:fits}}
\end{center}
\end{table}
\def\arraystretch{1}  

\begin{figure}[h,t,b] 
\begin{center}
\mbox{\epsfig{file=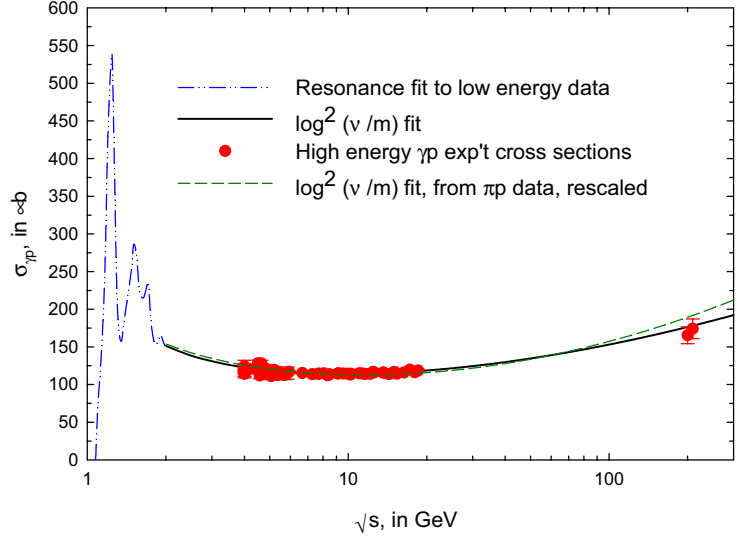,width=4.4in%
            ,
		bbllx=60pt,bblly=250pt,bburx=530pt,bbury=585pt,clip=%
}}
\end{center}
\caption[]{ \footnotesize
The dashed-dot-dot line is a fit by Damshek and Gilman\cite{gilman} of the low energy $\sigma_{\gamma p}$ data to a sum of five Breit-Wigner resonances plus a sixth-order polynomial background. The fit labeled ${\rm log}^2 (\nu/m)$, the solid line, is a $\chi^2$ fit (Fit~1) of the high energy data  of the form $\sigma_{\gamma p}=c_0 +c_1{\ln }(\nu/m)+c_2{\ln }^2(\nu/m)+\beta_{\cal P'}/\sqrt{\nu/m}$, with $\beta_{\cal P'}=$ 64 $ \mu$b.   
The laboratory energy of the photon  is  $\nu$ and $m$ is the proton mass. The dashed curve is from a $\sigma_{\pi p}=c_0 +c_1{\ln }(\nu/m)+c_2{\ln }^2(\nu/m)+\beta_{\cal P'}/\sqrt{\nu/m}$ fit from Igi and Ishida\cite{igi}, using their $\pi p$ cross sections rescaled by a factor of 1/208.  The cross sections join at $\sqrt s=2.01$ GeV, where their value is 151 $\mu$b.
}
\label{fig:siggpallenergies}
\end{figure}

In a recent paper, Igi and Ishida\cite{igi} have analyzed $\pi^+p$ and $\pi^-p$ total cross sections using parametrizations of the form of \eq{sigma}, with both $\ln(\nu/m)$ and $\ln^2(\nu/m)$. They derive two finite energy sum rules, which, when applied to the low energy data give two constraint equations.  Subject to these constraints, they make asymptotic fits to  the available high energy  $\pi^+p$ and $\pi^-p$ data.  Their analysis also favors a $\ln^2(\nu/m)$ behavior of the cross sections.  In a vector meson dominance model, the cross section $\sigma_{\gamma p}$ is proportional to the crossing-even $\pi p$ cross section, $\sigma_{\pi p}=(\sigma_{\pi^+p}+\sigma_{\pi^-p})/2$. We can thus confront our results above with  $\sigma_{\pi p}$ data, renormalized by a factor 208 familiar from vector meson dominance phenomenology. The result is shown as the dashed line in Fig.\,\ref{fig:siggpallenergies}. The agreement is excellent over a large energy interval---particularly, since the pion fit was derived using data only up to $\sqrt s\approx 30$ GeV , whereas the plot in Fig.\,\ref{fig:siggpallenergies} extends to 300~GeV.

\begin{figure}[h,t,b] 
\begin{center}
\mbox{\epsfig{file=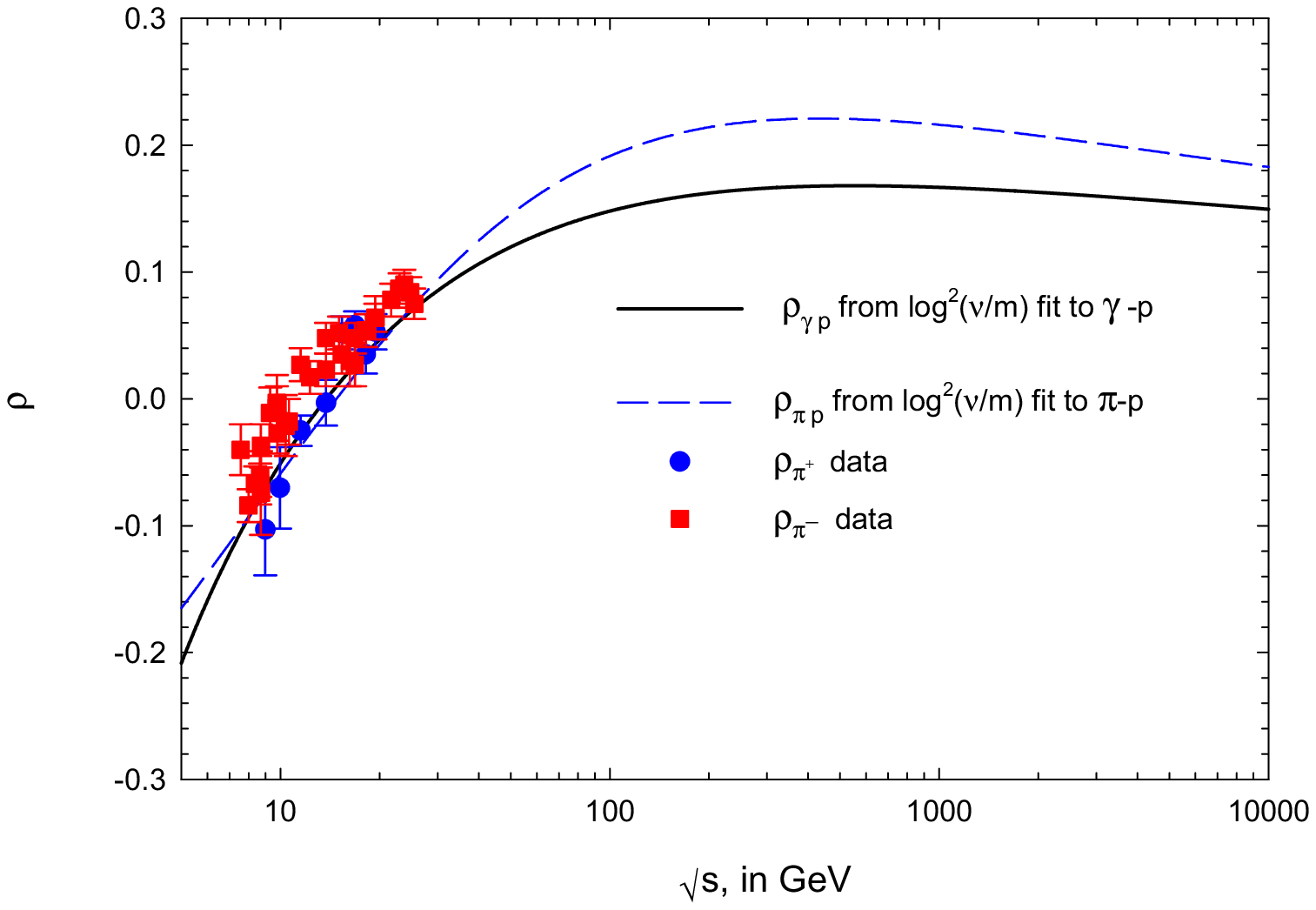,width=4.4in%
            ,
		bbllx=60pt,bblly=250pt,bburx=530pt,bbury=585pt,clip=%
}}
\end{center}
\caption[]{ \footnotesize
A plot of the real-to-imaginary ratio $\rho $ {\em vs.}$\sqrt s$, in GeV. The solid curve is  $\rho_{\gamma p}$, calculated from the parameters of Fit~1. The dashed curve is $\rho_{\pi p}$, calculated from a $\log^2{\nu}$ fit to $\pi^+ p$ and $\pi^- p$ data\cite{igi}.  The data shown are the experimental $\rho$ values\cite{pdg} for $\pi^+ p$ and $\pi^- p$.
}
\label{fig:rho}
\end{figure}

Finally, Block and Kaidalov\cite{bk} have suggested that factorization requires $\rho_{nn}=\rho_{\gamma p}$, and by extension,  $\rho_{\gamma p}$ should be equal to $\rho_{\pi p}$, where $\rho_{\pi p}$ is the even amplitude for $\pi^+p$ and $\pi^-p$ scattering. To test this, we plot $\rho_{\gamma p}$ as the solid curve in Fig.\,\ref{fig:rho}, using the parameters of Fit~1 in \eq{rho}. To calculate $\rho_{\pi p}$, we have taken the $\pi p$ fit parameters from Igi and Ishida\cite{igi} and have substituted them in \eq{rho}, where we have also set $f_+(0)=0$.  The  dashed curve in Fig.\,\ref{fig:rho} is $\rho_{\pi p}$. Also shown in Fig.\,\ref{fig:rho} are the high energy experimental $\rho$ data for $\pi^+p$ (the circles) and $\pi^-p$ (the squares). We note that $\rho_{\pi p}=(\rho_{\pi+ p}+\rho_{\pi^- p})/2$. The curves for 
$\pi p$ and for $\gamma p$ both reasonably describe the data, and essentially agree in the energy region below $30$ GeV---again, we emphasize that the $\pi p$ fit was only made for energies below 30 GeV. 

In conclusion, we have demonstrated that the duality requirement that high energy cross sections smoothly interpolate into the resonance region strongly favors a $\ln^2s$ behavior of the asymptotic cross section---a behavior that saturates the Froissart bound.  Using vector meson dominance we demonstrate that our conclusions are  also supported by $\pi^+p$ and $\pi^-p$ data. Our analysis supports predictions of Large Hadron Collider cross sections that rely on the saturation of the Froissart bound.

\vspace{.25in}

We thank Martin Olsson for suggesting the advantage of doing the analysis with photoproduction data. The work of FH is supported  in part by the U.S.~Department of Energy under Grant No.~DE-FG02-95ER40896 and in part by the University of Wisconsin Research Committee with funds granted by the Wisconsin Alumni Research Foundation.

%

%

\begin{thebibliography}{99}
\bibitem{sensitivity}Numerical calculations have shown that our results are insensitive to the choice of $\mu$ in \eq{sigma} and \eq{rho}, as well as the magnitude and slope of the cross section at the joining point, $\sqrt s$=2.01 GeV.
\end{thebibliography}
\end{document}